\documentclass[ 12 pt]{article}
\usepackage[a4paper, total={6in, 8in}]{geometry}
\usepackage{graphicx} 
\usepackage{amsmath}
\usepackage{graphicx}
\usepackage{cancel}
\usepackage{lipsum}
\usepackage{cite}
\usepackage{physics}
\usepackage{amsmath,amssymb}
\usepackage{dirtytalk}
\usepackage{xcolor}

\newcommand{\ed}{\end{document}}
\newcommand{\beq}{\begin{equation}}
\newcommand{\eeq}{\end{equation}}
\newcommand{\beqa}{\begin{eqnarray}}
\newcommand{\eeqa}{\end{eqnarray}}
\newcommand{\bc}{\begin{center}}
\newcommand{\ec}{\end{center}}

\newcommand{\ba}{\begin{array}}
\newcommand{\ea}{\end{array}}

\begin{document}
%\begin{widetext}
%\begin{widetext}
\title{Circuit Quantisation  in Hamiltonian Framework:\\A  Constraint Analysis Approach }% Force line breaks with \\

\author{Akshat Pandey\footnote{apandey.physics@gmail.com}  and Subir Ghosh\footnote{subirghosh20@gmail.com}   \\ \normalsize Physics and Applied Mathematics Unit\\Indian Statistical Institute, \\ 203, Barrackpore Trunk Road, Kolkata 700108, India} 
\date{}

 \maketitle

\begin{abstract}
   In this work  we apply Dirac's Constraint Analysis (DCA) to solve Superconducting Quantum Circuits (SQC). The Lagrangian of a SQC reveals the constraints, that are classified in a Hamiltonian framework, such that redundant variables can be removed to  isolate the canonical degrees of freedom for subsequent quantization of the Dirac Brackets. We demonstrate the robustness of DCA unlike certain other set of ideas like null vector and loop charge which are each applicable only to specific types of quantum circuits. 
\end{abstract}

\section{Introduction}

Quantum electromagnetic Circuits \cite{nn1} are ubiquitous in the major operational models of Quantum computing. Although they are analogous to classical computers in the sense that there are wires connected to gates that manipulate (qu)bits, however unlike their classical counterparts these transformations are completely reversible.  In particular, SQCs have become essential for  Quantum Computation  \cite{qcomp} with the designing of qubits being extended to include quantronium
\cite{quant}, transmon \cite{trans}, fluxonium \cite{flux}, and
“hybrid” \cite{hybrid} qubits, etc. At the macroscopic level, quantum effects are manifested in the field of circuit quantum electrodynamics \cite{pr87, sqed} such that SQCs as qubits  interact strongly in a controlled way  with microwave photons.

The question that we address in the present paper is the following \textbf{---} given a superconducting circuit, how can the Canonical Degrees of Freedom (CDOF) be identified in general. This identification which is essential for quantisation of the circuits continues to be an open problem \cite{sym}.  Different types of SQCs elements like Josephson Junctions and Quantum Phase Slips seem to require different methods - \textit{eg} graph theory  \cite{sym} or loop charges  \cite{dual}. In this current work, we demonstrate how Dirac's constraint analysis framework offers a simple and robust way to identify CDOFs in these circuits and to quantise them. The method, in addition to being generalisable to different types of circuits, offers additional insight into the freedom and choice of the CDOFs.

Before moving on, we would like to address a few subtle issues. SQCs, analogous to microscopic qubits,  are constructed out of macroscopic entities such as  electrical (LC) oscillator.  the collective motion of electron passing  without friction  at low temperature  and  Josephson effect,  introducing   nonlinearity without  dissipation or dephasing,
 is described by  flux threading the inductor,  acting as  center-of-mass position in a mass-spring mechanical oscillator. These conventional inductor-capacitor-resistance (LCR) circuits with batteries have a traditional Lagrangian description in terms of magnetic fluxes and electric charges. In SQC theory a Hamiltonian framework is favoured since it  yields  quantum commutators from classical Poisson brackets via Correspondence Principle. CDOFs are essential for the latter, and the presence of constraints makes DCA essential for a correct Hamiltonian description. As mentioned above, in addition to an unambiguous identification of CDOFs, DCA has the additional advantage that it can reveal additional symmetries of the circuit that within this formalism would manifest as gauge invariance. This can be used  to construct different circuits that are physically equivalent. Further, DCA lets us build circuits with preferred sets of CDOF, see section 3.2. Lastly, we also note that Symplectic Geometry formulation has also been used \cite{sym1,sym}, but possibly it is more convenient and economical to follow the Faddeev-Jackiw symplectic framework \cite{fad}.

This paper is organised as follows. In section 2 we briefly sketch the formalism of Dirac's Constraint Analysis (DCA). In section 3, we apply DCA to solve for, first a specifc SQC and then a generic SQC. We shall then discuss the quantisation of these circuits following DCA. We end with a summary in section 4.

\section{Constraint analysis in a nutshell}
We quickly recapitulate the salient features, (that will be particularly used here), of Dirac's Hamiltonian formulation of constraint analysis.

For a generic Lagrangian $L(q_i,\dot{q}_i)$, canonical momenta are defined as
\begin{equation}
p_i=(\partial L)/(\partial \dot{q}_i)
 \label{d1}   
\end{equation}
that satisfy canonical Poisson Brackets (PB) $\{q_i,p_j\}=\delta_{ij},~\{q_i,q_j\}=\{p_i,p_j\}=0 $.
The equations of the form
\begin{equation}
\chi_k=p_k-f_k(q_j)\approx 0
 \label{d2}   
\end{equation}
that do not contain $\dot{q}_i$ are treated as constraints. Next the extended form Hamiltonian, taking into account the constraints, is given by 
\begin{equation}
H_E=p_i\dot{q}_i-L+\alpha_k\chi_k 
 \label{d3}   
\end{equation}
where $\alpha _k$ are Lagrange arbitrary multipliers.
Time persistence of $\chi_k$, given by 
\begin{equation}
\dot \chi_k=\{\chi_k,H_E\}\approx 0
 \label{d4}   
\end{equation}
can give rise to further new constraints or in some cases some of the $\alpha_k$ can be determined (thus not generating new constraints). Note that "$\approx$" is a weak equality, modulo constraints, and  can not be used directly, in contrast to "$=$" strong equality (that appears later) that can be used directly on dynamical variables. This process will continue until no further constraints are generated. Once the full set of linearly independent constraints are determined the First Class Constraints (FCC), (associated with gauge invariance),  $\psi_r$ obeying $\{\psi_r,\chi_k\}\approx 0$ for all $k$ are isolated. Rest of the constraints are named Second Class Constraints (SCC). These are used to compute the all important non-singular constraint matrix $\chi_{kl}=\{\chi_k,\chi_l\}$. The SCCs can be used as strong relations to eliminate some of the DOF provided  Dirac brackets (DB), for two generic variables $A,B$, are used,
\begin{equation}
\{A,B\}_{DB}=\{A,B\}-\{A,\chi_r\}{\chi}_{rs}^{-1}\{\chi_s,B\},
 \label{d5}   
\end{equation}
where ${\chi}_{rs}^{-1}{\chi}_{st}=\delta_{rt}$. Notice that $\{A,\chi_r\}_{DB}=0$ for any $A$ which justifies using $\chi_r=0$ so that each SCC $\chi_r$  removes one  DOF. Thus finally one has the Hamiltonian comprising of a reduced number of variables and time evolution of $A$ is computed using DBs
\begin{equation}
\dot A=\{A,H\}_{DB}.
 \label{df4}   
\end{equation}
Notice that, under normal circumstances, there will be an even number of SCCs since the matrices $\chi$ and $\chi^{-1}$ are anti-symmetric and non-singular.

One still has the FCC which reveals gauge symmetry of the model. The FCCs,  together with respective gauge fixing conditions, give rise to a further set of SCCs, further  DBs, that will further  reduce the number of DOF. It is important to note that each SCC can remove one DOF in phase space and each FCC can remove two DOF in phase space (because of its associated gauge condition) leading to the formula

\begin{align*}
({\text{number of physical DOF in phase space}})= ({\text{total number of DOF in phase space}}) \\  
-(2\times ({\text{number of FCC}})+1\times ({\text{number of SCC}})).
 \label{dff4} 
\end{align*}

Due to the freedom of  choice in (allowed) gauge conditions there can be manifestly different but gauge equivalent models. 

\section{Solving SQC using DCA}

\subsection{DCA of Inductively Shunted SQC} 
\begin{figure}[htp]
    \centering
    \includegraphics[width=4cm]{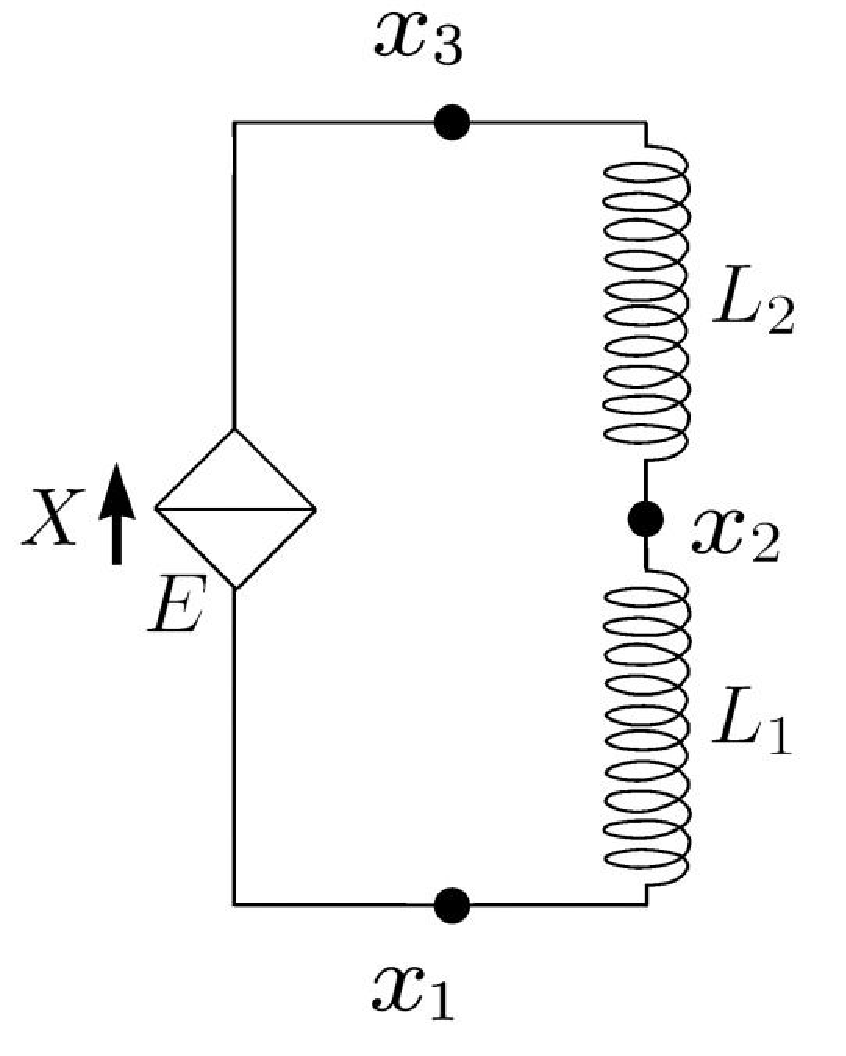}
    \caption{A circuit involving an inductively shunted island at $x_{2}$ \cite{sym}. }
    \label{fig1}
\end{figure}

We borrow this model from \cite{sym} with  corresponding  SQC in Figure (\ref{fig1}), 

\begin{equation}
L=X(\dot{x}_3-\dot{x}_1) +Ecos(\frac{2\pi X}{2e})-\frac{1}{2L_1}(x_1-x_2)^2 -\frac{1}{2L_2}(x_2-x_3)^2
    \label{1}
\end{equation}
In the terminology of \cite{sym} $x_i$ are three flux node variables and $X$ is the branch charge.
\begin{equation}
P_i=\frac{\partial L}{\partial \dot{x}_i}~,~i=1,2,3,~~ \pi =\frac{\partial L}{\partial \dot{X}}~;~\{x_i,P_{j}\}=\delta_{ij},~~\{X,\pi\}=1.
    \label{2}
\end{equation}
 The constraints are $\chi_1=P_1+X\approx 0,~\chi_2=P_2\approx 0,~\chi_3=P_3-X\approx 0,~\chi_4=\pi\approx 0$, but we use an equivalent linear combination 
\begin{equation}
\chi_1=P_1+P_3\approx 0,~\chi_2=P_2\approx 0,~\chi_3=P_3-X\approx 0,~\chi_4=\pi\approx 0 .
    \label{3a}
\end{equation}
The extended Hamiltonian is
\begin{equation}
H_E=-Ecos(\frac{2\pi P_3}{2e})+\frac{1}{2L_1}(x_1-x_2)^2 +\frac{1}{2L_2}(x_2-x_3)^2 +\sum_{i=1}^4 \alpha_i\chi_i .
    \label{5}
\end{equation}
 Time persistence $\dot{\chi_i}=\{\chi_i,H_{E}\}\approx 0$ reproduces  $\alpha_3,\alpha_4$ from $\dot{\chi_3}=0, \dot{\chi_4}=0 $ respectively. Since the SCC pair $\chi_3,\chi_4$ commutes with the other constraints,   DBs generated by them  can be calculated in the first stage - a useful property of DCA. In the present case, this does not induce any change in the brackets of the variables of interest.

Time persistence of the remaining constraints does not determine   $\alpha_1,\alpha_2$ but  generates a new constraint 
\begin{equation}
\dot {\chi}_1=\{\chi_1,H\}=((L_1+L_2)x_2-(L_1x_3+L_2x_1))/(L_1L_2)\equiv \chi_5
\approx 0, ~ \dot {\chi}_2=-\equiv \chi_5\approx 0 .
    \label{8}
\end{equation}
It can be shown  that $\Psi=\chi_1+\chi_2=P_1+P_2+P_3 \approx 0$ is a FCC since it commutes with rest of the constraints $\chi_1,\chi_2,\chi_5$. Since $\dot \Psi =\{\Psi,H\}\approx 0$ the chain of constraints stops and there are no further constraints and
\begin{equation}
H=-Ecos(\frac{2\pi P_3}{2e})+\frac{1}{2L_1}(x_1-x_2)^2 +\frac{1}{2L_2}(x_2-x_3)^2.
    \label{7}
\end{equation}
Since $\Psi$ is already considered as an FCC we can choose $\chi_5$ and either of $\chi_1,\chi_2$ as the SCC pair; we choose $\chi_2,\chi_5$ and with DBs in mind, use them strongly to obtain 
\begin{equation}
H=-Ecos(\frac{2\pi P_3}{2e})+\frac{1}{2(L_1+L_2)}(x_1-x_3)^2
    \label{10}
\end{equation}
The brackets between $P_3,x_1,x_3$ are not affected. Notice that $\{P_3,x_1-x_3\}=1$ is a canonical pair. This agrees with our DOF count since there are overall four SCC and one FCC $\rightarrow~ 8-(1\times 2+4\times 1)=2$ DOF in phase space or one DOF in configuration space. 

\begin{figure}[htp]
    \centering
    \includegraphics[width=4cm]{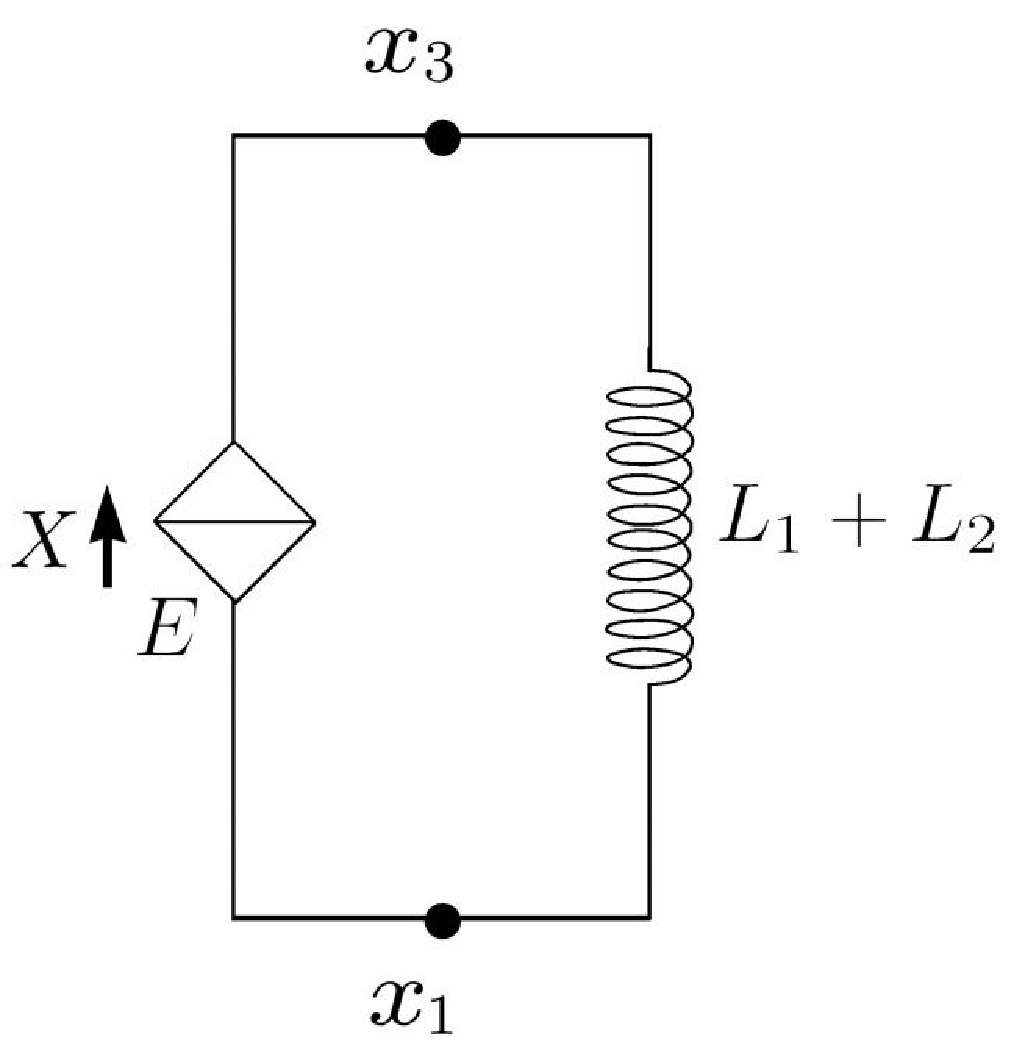}
    \caption{A dynamically equivalent circuit to figure 1 with the non-dynamical variables removed \cite{sym}. }
    \label{fig2}
\end{figure}

We still have the FCC $\psi\approx 0$ for which we can choose a gauge constraint $\Theta=x_1\approx 0$, (such that $(\psi,\Theta)$ are SCC), leading to
\begin{equation}
H=-Ecos(\frac{2\pi P_3}{2e})+\frac{1}{2(L_1+L_2)}(x_3)^2 .
    \label{11}
    \end{equation}
    Thus using DCA  we have recovered, much more economically, all the results of \cite{sym}, with an additional insight of a gauge invariance.

    As advertised earlier, we now show the flexibility induced by the FCC; a generalized  gauge condition $\Theta=ax_1+bx_3,~a,b \text{ arbitrary constants}$ will generate  different 
 DBs
\begin{equation}
\{x_1,P_1\}_{DB}=\frac{b}{a+b},~\{x_1,P_3\}_{DB}=-\frac{b}{a+b},$$$$
\{x_3,P_3\}_{DB}=\frac{a}{a+b},~\{x_3,P_1\}_{DB}=-\frac{a}{a+b} .
    \label{13}
    \end{equation}
    Using the gauge condition  $x_1=-\frac{b}{a}x_3$ we get $H$ and equations of motion
\begin{equation}
H=-Ecos(\frac{2\pi P_3}{2e})+\frac{1}{2(L_1+L_2)}(\frac{a+b}{a})^2(x_3)^2
    \label{14}
\end{equation}	
\begin{equation}
\dot{x}_3=\frac{a}{a+b}\frac{2\pi E}{2e}sin (\frac{2\pi P_3}{2e}),~\dot{P}_3=-\frac{a+b}{a}\frac{x_3}{L_1+L_2}
    \label{15}
\end{equation}	 
finally yielding
\begin{equation}
\ddot{x}_3=-(\frac{2\pi}{2e})^2 \frac{E}{L_1+L_2}cos (\frac{2\pi P_3}{2e})x_3.
    \label{16}
\end{equation}	
Note that (\ref{16}) it is independent of $a,b$ and so gauge invariant. Also notice that the bracket
$\{x_3,P_3\}_{DB}=\frac{a}{a+b}$ can be directly identified from the Lagrangian
\begin{equation}
L=\frac{a+b}{a}\dot{x}_3P_3 +Ecos(\frac{2\pi P_3}{2e})+\frac{1}{2(L_1+L_2)}(\frac{a+b}{a})^2(x_3)^2
    \label{ll}
\end{equation}	
and in fact with a redefinition $\frac{a+b}{a}x\rightarrow x$ the system becomes trivially canonical.

Here we would like to note that the in the above analysis, the details of the circuit components (inductors or capacitors) were not explicitly used, unlike in \cite{sym}. As a consequence, DCA turns out be a very general scheme to solve SQCs \cite{arxiv}. 

\subsection{DCA of a generic SQC} We shall now treat a generic SQC system, without going into the details what the circuit elements are. The purpose of this is to offer a general scheme which could be applied to various specific circuits. We study the following Lagrangian
\begin{equation}
L=X(\dot{x}_3-\dot{x}_1) +E  cos(\frac{2\pi X}{2e})-\frac{1}{2L_1}(x_1-x_2)^2 -\frac{1}{2L_2}(x_2-x_3)^2 +\lambda_1\dot{x}_1x_3 +\lambda_2\dot{x}_1x_2 +\lambda_3\dot{x}_2x_3 .
    \label{nn1}
\end{equation}
This Lagrangian, for example, allows for a circuit analogue to non-commutative QM \cite{arxiv, noncom}.

Here the constraints are constraints $
\chi_1=P_1+P_3-\lambda_1x_3-\lambda_2x_2\approx 0,~\chi_2=P_2-\lambda_3x_3\approx 0,~\chi_3=P_3-X\approx 0,~\chi_4=\pi\approx 0 $ and 
 extended Hamiltonian
\begin{equation}
H=-Ecos(\frac{2\pi P_3}{2e})+\frac{1}{2L_1}(x_1-x_2)^2 +\frac{1}{2L_2}(x_2-x_3)^2 +\sum_{i=1}^4 \alpha_i\chi_i
    \label{nn5}
\end{equation}
reveal  that  $\dot{\chi}_i\approx 0$ fixes all the $\alpha_i$, so there are no further constraints. The constraint matrix $\{\chi_i,\chi_j\}$ is non-singular; $\chi_i$s comprise a set of four SCC and  no FCC as in  previous cases. We will have $8-(4\times 1)=4$ DOF in phase space or two independent DOF in configuration space.

The full set of DBs among "coordinate" variables are given by 
\begin{equation}
\{x_1,x_2\}_{DB}=\frac{1}{\lambda_2-\lambda_3},~\{x_1,x_3\}_{DB}=0,~\{x_1,X\}_{DB}=\frac{\lambda_3}{\lambda_2-\lambda_3}, $$$$
\{x_2,x_3\}_{DB}=-\frac{1}{\lambda_2-\lambda_3},~\{x_2,X\}_{DB}=-\frac{\lambda_1}{\lambda_2-\lambda_3},~\{x_3,X\}_{DB}=\frac{\lambda_2}{\lambda_2-\lambda_3}
    \label{n6}
\end{equation}	 
 The significance of noncommutative quantum circuits is now clear since conventionally noncommutative quantum mechanics refers to models where spatial coordinates do not commute.  Rest of the DBs are can also be computed in a straightforward way but for the present model these are not required. We can exploit the SCCs $\chi_i=0$ as strong relations to express the Hamiltonian as
\begin{equation}
H=-Ecos(\frac{2\pi X}{2e})+\frac{1}{2L_1}(x_1-x_2)^2 +\frac{1}{2L_2}(x_2-x_3)^2 +\sum_{i=1}^4 \alpha_i\chi_i .
    \label{nn7}
\end{equation}
It is interesting to note that the under DB the DOFs $(Q_1,P_1)$ and $(Q_2,P_2)$  constitute {\it{two independent canonical degrees of freedom}} where 
\begin{equation}
 Q_1=(x_3-x_1),~P_1=X, ~Q_2=x_2+\frac{\lambda_1}{\lambda_3}x_1,~P_2=-\lambda_2x_1+\lambda_3x_3 ,
     \label{n8}
\end{equation}	
with the only non-vanishing DBs being $\{Q_1,P_1\}_{DB}=1,~\{Q_2,P_2\}_{DB}=1$. This result agrees with our advertised DOF count. 

Let us invert the relations (\ref{n8})
\begin{equation}
x_1=\frac{\lambda_3-P_2}{\lambda_2-\lambda_3},~x_2=\frac{Q_2\lambda_3(\lambda_2-\lambda_3)-\lambda_1(\lambda_3Q_1-P_2)}{\lambda_3(\lambda_2-\lambda_3)},~x_3=\frac{Q_1\lambda_2-P_2}{\lambda_2-\lambda_3},~X=P_1 .
    \label{n9}
\end{equation}	
Keeping the unique noncommutative feature of the circuit intact let us consider a simplified model with $\lambda_1=\lambda_2=0,~\lambda_3=\lambda$ so that the Dirac algebra simplifies to
\begin{equation}
\{x_1,x_2\}_{DB}=-\frac{1}{\lambda},~\{x_1,x_3\}_{DB}=0,~\{x_1,X\}_{DB}=-1,~
\{x_2,x_3\}_{DB}=\frac{1}{\lambda}, $$$$
\{x_2,X\}_{DB}=0,~\{x_3,X\}_{DB}=0 .
    \label{n6a}
\end{equation}	
The Hamiltonian  is expressed in terms of canonical variables as 
\begin{equation}
H=-Ecos(\frac{2\pi P_1}{2e})+\frac{1}{2L_1 \lambda^2}(\lambda (Q_1+Q_2)-P_2)^2 +\frac{1}{2L_2\lambda^2}(\lambda Q_2-P_2)^2  .
    \label{n10}
\end{equation}	 
This will yield the equations of motion
\begin{equation}
\ddot{Q}_1=(\frac{2\pi}{2e})^2Ecos(\frac{2\pi P_1}{2e})(-\frac{Q_1+Q_2}{L_1}+\frac{P_2}{\lambda L_1}), ~
\ddot{Q}_2=-\frac{1}{L_2\lambda^3}(\frac{\lambda}{L_1}((Q_1+Q_2)-P_2).
    \label{n11}
\end{equation}	
Notice that for a low momentum approximation $P_1\approx 0,~P_2\approx 0$, the system reduces to a single simple harmonic oscillator, (for $Q_1+Q_2=Q$),
\begin{equation}
\ddot Q=-\left  ((\frac{2\pi}{2e})^2\frac{E}{L_1}+\frac{1}{L_1L_2\lambda^3}\right )Q .
    \label{n12}
\end{equation}	

\subsection{Circuit Quantisation}
The true power of Dirac's formalism manifests itself in {\it{quantization of constraint systems}}. An essential reason for this is that Dirac's scheme is based on Hamiltonian formalism; a generalized (Poisson-like or Dirac) bracket structure that can be directly exploited in transition to quantum regime. In this scheme a generalization of the Correspondence Principle is proposed \cite{dir}; in case of constraint systems the Dirac Brackets have to be elevated to the status of quantum commutators,
\begin{equation}
    \{A, B\}_{DB}=C ~~\rightarrow ~~[\hat A,\hat B]=i\hbar \hat C
    \label{d1}
\end{equation}
 In some models (in particular where the constraints are non-linear functions of the dynamical variables), $C$ can depend on  the dynamical variables as well leading to operator valued $\hat C$ upon quantization. This can give rise to  complications but there are systematic ways of treating such problems. But in the SQCs considered in this work, that happens to be the generic cases in this context, $C$ in (\ref{d1}) is a constant (matrix) that can easily be scaled away to yield a canonical form.

\section{Summary}

%From surveying recent literature it appears that a unified framework is required in the  theoretical analysis  of Superconducting Quantum Circuits, a key element in quantum information and quantum computing technology. Qualitatively distinct schemes are exploited for different forms of SQCs (involving sophisticated mathematical tools) that generate correct results. Indeed it would be desirable if a systematic and conceptually clear formalism is found that can address all types of SQCs in equal footing and yield    the effective canonical degrees of freedom, that have to be quantized eventually. This issue is non-trivial because in the Lagrangian approach, complications can arise due to the presence of constraints. \\  
In the present work we have conclusively established that the Hamiltonian framework of constraint dynamics, as formulated by Dirac, is highly appropriate to analyse Superconducting Quantum Circuits. We solved a particular model of an SQC from the literature, namely one with an inductively shunted island, and showed how it leads to an unambiguous removal of non-dynamical variables irrespective of the details of the circuit element. We then carried out the same analysis for a generic SQC that would with different addition constraints correspond to various known SQC systems. We then discussed the quantisation of these systems and emphasised the universality of the DCA framework in such systems.

Finally we would like to comment on a recent work \cite{mart} that discusses "failure" of DCA in case of  nearly singular SQC; in particular examples of SQC with very low capacitance (effectively playing the role of a small pass particle in a mechanical circuit) are suggested. It needs to be appreciated that the difference between very small "mass" and exactly zero "mass" can be critical as it can very well introduce (or remove as the case may be) symmetries of the system, that will be reflected in the nature of constraint classification, changing Second Class to First Class Constraint (or {\it{vice versa}}). For example in (\ref{d5}), if the constraint matrix $\chi_{kl}=function ~of ~"mass"$, it might so happen that the RHS vanishes for $"mass"=0$ so that $\chi_{kl}$ becomes singular, indicating that some of the Second Class Constraints are now First Class, leading to additional symmetries (and maybe new degrees of freedom). Obviously one will get nonsensical results if one uses DCA and tries to  obtain $"mass"=0$ results by considering the zero "mass" limit of $\chi_{kl}=function ~of ~"mass"$. DCA demands, in this case, that  non-zero "mass" and zero "mass" models are to be treated individually, and not as a zero "mass" limit from former to latter model. Furthermore, it needs to be stressed that in \cite{mart} problems appear when the system Hamiltonian becomes multiple valued such that the model itself is pathological; the failure is not a weakness of DCA.   Hence  the claim is that given the Lagrangian for any SQC, however complicated, the method presented here, can be exploited in solving the circuit dynamics. On the other hand, going in the reverse direction, one can posit a Lagrangian with certain preconceived featured, solve the model in Dirac framework and subsequently visualize the corresponding SQC. We would also like to mention an alternative way to identify and handle pathological models, which we believe deserves furhter investigation; this is, namely, the use of "Feynman rules for quantum circuits" introduced by Srivastava and Widom\cite{pr87}. These rules are formulated from the bottom up and thus studying SQCs in this framework can help understand the underlying physical origin of these pathologies and explain why they can be neglected.\\ 

\section*{Funding Statement}

No funding was involved in the completion of this project.

\section*{Contributions}

SG conceptualised the problem. In carrying out the detailed calculations, analysis, writing and editing both AP and SG contributed equally.

\section*{Conflict of Interest Statement}

The authors declare no conflict of interest.

\end{document}